\documentclass[%
 reprint,
showpacs,preprintnumbers,
 amsmath,amssymb,
 aps,
pra,
]{revtex4-1}

\usepackage{graphicx}
\usepackage{dcolumn}
\usepackage{bm}

\usepackage{graphicx}
\usepackage{color,epsfig}
\usepackage{bm}
\usepackage{amsmath,amsfonts,amssymb,bm}
\usepackage{natbib}

\usepackage[colorlinks=true,%
            linkcolor=blue,%
            urlcolor=blue,%
            citecolor=blue,%
            filecolor=blue,%
            bookmarksopen=true,%
            pdfauthor={},%
            pdftitle={Quantum Otto engine for a graphene quantum dots},%
            pdfsubject={},%
            pdfpagemode=UseOutlines]{hyperref}

\begin{document}

\title{Quasi-static and quantum-adiabatic Otto engine for a 2-D material: the case of a graphene quantum dot}

\author{Francisco J. Pe\~na}
\email[F. J. Pe\~na]{\quad francisco.penar@usm.cl}
\affiliation{Departamento de F\'isica, Universidad T\'ecnica Federico Santa Mar\'ia,  2390123 Valpara\'iso, Chile.}

\author{D. Zambrano}
\affiliation{Departamento de F\'isica, Universidad T\'ecnica Federico Santa Mar\'ia,  2390123 Valpara\'iso, Chile.}
 
\author{O. Negrete}
\affiliation{Departamento de F\'isica, Universidad T\'ecnica Federico Santa Mar\'ia,  2390123 Valpara\'iso, Chile.\\
		Centro para el Desarrollo de la Nanociencia y la Nanotecnolog\'ia, 8320000 Santiago, Chile.}

\author{Gabriele De Chiara}
\affiliation{Centre for Theoretical Atomic, Molecular and Optical Physics,
School of Mathematics and Physics, Queen's University Belfast, Belfast BT7 1NN, United Kingdom.}

\author{P. A. Orellana}
\affiliation{Departamento de F\'isica, Universidad T\'ecnica Federico Santa Mar\'ia,  2390123 Valpara\'iso, Chile.}

\author{P. Vargas}
\affiliation{Departamento de F\'isica, Universidad T\'ecnica Federico Santa Mar\'ia,  2390123 Valpara\'iso, Chile.\\
		Centro para el Desarrollo de la Nanociencia y la Nanotecnolog\'ia, 8320000 Santiago, Chile.}

\date{\today}

\begin{abstract}
In this work, we study the performance of a quasi-static  and quantum-adiabatic magnetic Otto cycles with a working substance composed of a single graphene quantum dot modeled by the continuum approach with the use of the zigzag boundary condition. Modulating an external/perpendicular magnetic field, in the quasi-static approach, we found a constant behavior in the total work extracted that is not present in the quantum-adiabatic  formulation. We find that, in the quasi-static  approach, the engine yielded a greater performance in terms of total work extracted and efficiency as compared with its quantum-adiabatic  counterpart. In the quasi-static case, this is due to the working substance being in thermal equilibrium at each point of the cycle, maximizing the energy extracted in the adiabatic strokes.
\end{abstract}

\maketitle


\section{\label{sec:level1}Introduction}

The concept of quantum heat engines (QHEs) was introduced by Scovil and Schultz-Dubois in \cite{Scovil}, in which they demonstrate that a three-level energy maser can be described as a heat engine operating under a Carnot cycle. This important research gave way to the study of quantum systems implemented as the working substances of heat machines with the goal of realizing efficient nanoscale devices. These devices are characterized by the structure of their working substance, the thermodynamic cycle of operation, and the dynamics that govern the cycle \cite{Bender_02,Feldmann,Feldmann2,Rezek1,Henrich,Huang_013,Li_Negentropy_013,  
Lutz_014,Scully_03,2Wang2012,Scully_011,munozpena2015,Wang_011,Abe1,Wang2012(1),Ruiwang2012,  
  H.Wang2013,Guo1,Hewgill2018,Chuankun1,Esposito1,munozpena2012,Abe2,Abe3,Wang2015, Zhang2014,delcampo2014,Fialko2012,Newman2017,Reid2018,Deffner1,Mitchison,Vischi,Vischi2}.  A QHEs cycle consists of a combination of quantum thermodynamics processes such as the quantum-adiabatic  process, the quantum isothermal process, the quantum isobaric process, and the quantum isochoric process \cite{Deffner2}. Therefore, we always have a quantum-adiabatic  version of the most famous cycles like Carnot, Ericsson, Brayton, and Otto. In particular, the quantum Otto cycle has been considered for different working substances such as spin-1/2 systems \cite{Huang2014,Sun2016}, harmonic oscillators \cite{Kosloff2017}, among others \cite{Hubner2014,Mehta2017,Munoz2014,Karimi2016,Zheng2014,ObinnaAbah2016,Zheng2016,cmk2016}. Furthermore, it has been shown that thermal machines can be reduced to the limits of single atoms \cite{Rossnagel2016}, and recently, the effects of the wave function symmetry on Otto's engine performance have been studied analytically \cite{Myers}, further increasing the interest of this incipient area.

On the other hand, quantum dots today have a robust architecture of devices based on them. There is always a search in the control of its size, shape, and distribution to characterize its optoelectronic properties in order to find future technological applications \cite{Devrim2014}. In this context, the case of quantum dots of GaAs or (InAs) under a controllable external magnetic field as a working substance operating under an Otto cycle has been studied recently \cite{Pena2019}, where the comparison has been made regarding the application of the quasi-static and quantum-adiabatic  performance of a multi-level Otto cycle in a diagonal formulation of the density matrix operator.

A possible extension of the work \cite{Pena2019} focused on the so-called 2-D materials \cite{Suarez2019}. To date, the most characterized and studied one is graphene. Graphene is a one-atom-thick covalently-bonded carbon layer ordered in a honeycomb lattice and has attracted considerable attention \cite{Castro2009}. One of the factors which makes graphene so attractive for research is the ultrafast low-energy dynamics of its charge carriers. Those carriers can be described by a two-dimensional Dirac-Weyl equation and linear dispersion relation.  For graphene quantum dots \cite{Pomarenko2008,Schnez2009}, the low energy approach using the Dirac equation with boundary conditions is an excellent approximation. We remark two approaches, the zig-zag boundary conditions, and infinite mass boundary conditions. The first one is related to the vanishing of one component of the spinor at the dot edge and the second one requires that the region outside the dot is forbidden for particles due to the relationship of the Fermi velocity in the form of $v_{f}\propto 1/m$. These two approaches satisfy the condition of zero current at the edge of the graphene dot \cite{Schnez2008,Schnez2017,Grujic2011,Thomsen2017}.

In this work, we study the performance of a quasi-static and quantum-adiabatic  Otto cycle in a diagonal formulation of the density matrix operator, where the working substance involves a graphene quantum dot under perpendicular external magnetic field. This system is described by  the using the continuum approach  (Dirac equation with boundary conditions) fully addressed by Gruji\'{c} \textit{et al.} \cite{Grujic2011} and recently extended for rings and antidots structures by Thomsen and Pedersen \cite{Thomsen2017}. We report that in the quasi-static approach the total work extracted is greater than its quantum-adiabatic  counterpart for high temperature behaviour while for low temperatures behaviour both cases studied tend to converge. In addition, for the quasi-static case, we find a region of parameters in which the total work extracted becomes independent of the change in the external parameter that governs the cycle, an effect which is not perceptible under the quantum-adiabatic  formulation.

\section{\label{sec:level2} Model}

We consider the Dirac-Weyl Hamiltonian for low energy electron states in graphene under the presence of external perpendicular magnetic field and a mass related potential given by 
\begin{equation}
	H=v_{F}\left(\textbf{p}+e\textbf{A}\right)\cdot \boldsymbol{\sigma} + V(r) \sigma_{z}\text{,}
    \label{dirachamil}
\end{equation}
where $v_{F}\sim 10^{6} m/s$, $\textbf{A}$ is the vector potential and $\bf{\sigma}=\left(\sigma_{x},\sigma_{y}\right)$ are Pauli's spin matrices. Eq. (\ref{dirachamil}) is valid for the $K$ valley states in graphene \cite{Grujic2011}. For the study of $K'$ valley states it is necessary to replace $\boldsymbol{\sigma}$ for its complex conjugate $\boldsymbol{\sigma}^{*}$. We take the model treated in the Refs. \cite{Grujic2011,Thomsen2017} where the authors assume that the carriers are confined to a circular area of radius $R$, which is modeled by a potential of the form
\begin{equation}
  V(r) = 
  \begin{cases} 
    0      & \mbox{if } r<R\text{,} \\
    \infty & \mbox{if } r\geq R\text{,} \\
  \end{cases}
  \label{potential}
\end{equation}
where $r$ is the radial coordinate of the cylindrical coordinates. There are two different boundary conditions that can be applied to treat the potential form of Eq. (\ref{potential}): the zigzag boundary conditions (ZZBC) and the  infinite mass boundary conditions (IMBC). For the case of ZZBC the two Dirac cones are labeled with the quantum number $k$, which has the value $+1$ in the $K$ valley and $-1$ in the $K'$ valley. For the IMBC however, the so-called valley-isotropic form of the Hamiltonian is used and the valleys are differentiated by another quantum number $\tau$ that appears in the IMBC formulation as a multiplicative factor to the potential $V(r)$ in Eq. (\ref{dirachamil}). First, we  will compare these two approximations used in the continuum approach and we will discuss why the selection of one over the other in the thermodynamic study of this work.

In order to obtain the energy spectrum of the graphene quantum dot previously reported \cite{Grujic2011,Thomsen2017}, we introduce the dimensionless variables $\rho=r/R$, $\beta=R^{2}/2l_{B}^{2}=eBR^{2}/2\hbar$ and $\varepsilon=E/E_{0}=ER/\hbar v_{f}$, where $E$ is the carrier energy and $l_{B}=\sqrt{\hbar/(e B)}$ is the magnetic length. It is very well known that the total angular momentum, $J_{z}$, containing the contributions of orbital angular momentum $(L_{z})$ and pseudospin ($\hbar\sigma_{z}/2$), commutes with the Hamiltonian of Eq. (\ref{dirachamil}) and is therefore a conserved quantity. Under these assumptions the two-component wave function must have the form 
\begin{align}
    \Psi(\rho,\phi)=\begin{pmatrix}
           \psi_{1}(\rho,\phi)\\           
           \psi_{2}(\rho,\phi) 
    \end{pmatrix}=
       e^{im\phi} \begin{pmatrix}
           \chi_{1}(\rho)\\           
           e^{ik\phi}\chi_{2}(\rho) 
    \end{pmatrix},
  \end{align}
where $m = 0, \pm 1, \pm $, ... is the total angular momentum quantum number and $\phi$ is the polar angle.

For the case of  IMBC, the charge carriers are confined inside the quantum dot. This leads to the infinite-mass boundary which yields the following condition between the components of the spinor:

\begin{equation}
	\frac{\psi_{1}\left(\rho^{*},\phi\right)}{\psi_{2}\left(\rho^{*},\phi\right)}=i\tau e^{i\phi}\text{,}
	\label{fundamental1}
\end{equation}

where $\rho^{*}$ correspond to the radial coordinate evaluate at the boundary ($r=R$, i. e. $\rho^{*}=1$). The solution of the time independent Dirac equation given by $H \Psi(\rho,\phi) = E \Psi(\rho,\phi)$ is fully addressed by Gruji\'{c} \textit{et al.} \cite{Grujic2011} and for the case of nonzero energy solutions and $\beta\neq 0$ (nonzero external field), IMBC leads to the following eigenvalue equation

\begin{eqnarray}
\frac{\tau\varepsilon}{2}&{_1 {{\widetilde{F}}}_{1}}&\left(m+1-\frac{\varepsilon^{2}}{4\beta}, m+2, \beta \right) \\ \nonumber 
- &{_1 {{\widetilde{F}}}_{1}}&\left(m+1-\frac{\varepsilon^{2}}{4\beta}, m+1, \beta \right)=0,
\end{eqnarray}
where ${_1 {{\widetilde{F}}}_{1}(a,b,z)}$ is the regularized confluent hypergeometric function.

On the other hand, ZZBC requires that one of the components of the spinor to vanish  at the boundary, that is

\begin{equation}
    \psi(\rho^{*},\phi)=0 \rightarrow \chi_{1}(\rho^{*})=0.
    \label{ZZBCcondition}
\end{equation}

The treatment of Dirac equation with the combination of Eq. (\ref{ZZBCcondition}) leads to an equation of eigenvalues of the form

\begin{eqnarray}
{_1 {{\widetilde{F}}}_{1}}\left(m+\frac{1}{2}+\frac{k}{2}-\frac{\varepsilon^{2}}{4\beta}, m+1, \beta \right)=0.
\end{eqnarray}


\begin{figure}[!ht]
\centering
\includegraphics[width=1.0\linewidth]{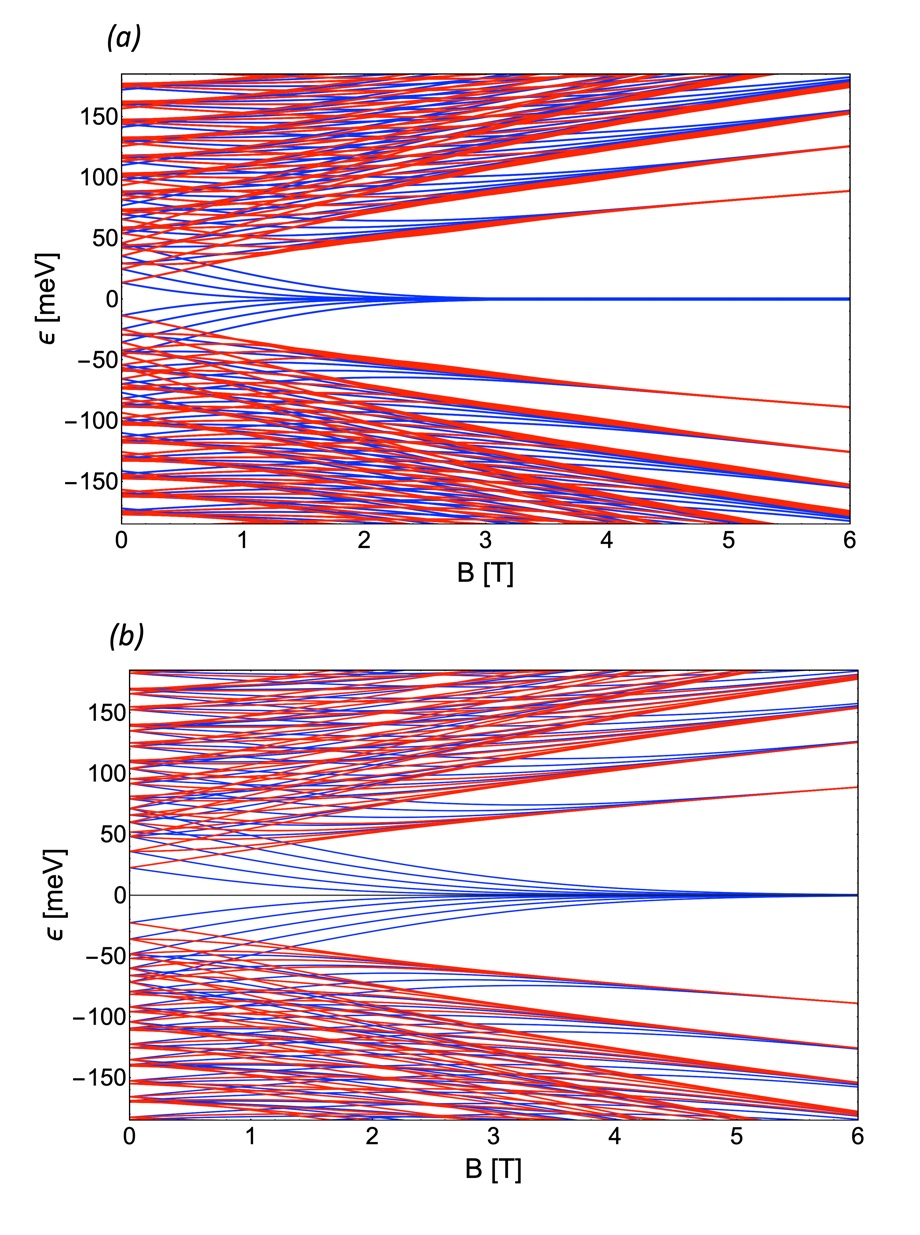}
\caption{Energy spectrum (in meV) of the graphene quantum dot of $R=70 \ nm $ as a function of the external magnetic field $B$ (in Tesla) for (a) the infinite mass boundary condition (IMBC) and (b) zigzag boundary condition (ZZBC). Only the six lowest electron and hole energy levels are shown for the azimuthal quantum number beteween $m=-4,...,0,...,4$. The red lines represent the solutions for $\tau=+1$ (or $k=+1$) and the blue lines the energy for $\tau=-1$ (or $k=-1$). In the (b) panel, the black line represent the zero energy solution.}
\label{fig_spectrum}
\end{figure}

The energy spectrum for a graphene quantum dot of $R= 70$ nm is presented in Fig. \ref{fig_spectrum} for a range of energy between $-200 $ meV and $200$ meV as a function of the perpendicular external magnetic field for IMBC ( (a) panel) and ZZBC ( (b) panel). A crucial difference between the two approaches is the presence of the zero-energy eigenstate for the ZZBC which is instead missing for the IMBC. For our model, the zero-energy state will be considered due to its importance confirmed in recent experiments \cite{Magda2014} and to the fact that in the case of IMBC it does not appear due to mathematical reasons.

In the work of Gruji\'{c} \textit{et al.} \cite{Grujic2011}, the authors discuss the influence of the boundary conditions on the energy spectra. In that work, the tight-binding approximation (TB) is compared with the continuum approach with ZZBC (as we use here for our QHE) and IMBC . They propose a model for TB of a circular region of graphene surrounded by an infinite-mass media and find that the continuum model with ZZBC converges very well for larger dots (i.e. $R > 10$ nm) and lower-energy states between $0$ eV to  $0.20$ eV approximately because some curves in the energy spectrum as function of the dot radius ($R$) obtained using the TB approximation do not decay monotonically as $\propto 1/R$ and exhibit some fluctuation behavior, which is more pronounced for smaller radii. In particular, they conclude that the microscopic details become important as $R$ decrease and cannot be described by the continuum approach. Therefore, it is important to recall that our working substance satisfies the conditions of large dot radius (i. e. we work for $R> 10$ nm ) and low energy spectra (i. e. we work between $0 $ meV to $200$  meV) 
so that the edge imperfections are less important.   

\section{THERMODYNAMICS QUANTITIES}
\label{thermoqua}
In order to obtain the thermodynamics of the system, we calculate the canonical partition function given by 
\begin{eqnarray}
\mathcal{Z}(T,B)=\sum_{m,\tau}e^{-\frac{E_{m,\tau}}{k_{B}T}},
\end{eqnarray}
where $E_{m,\tau}$ correspond to the energy levels of particle-like solution of the Dirac equation (i.e $E_{m,\tau}\geq 0$) calculated for the same parameters of Fig. \ref{fig_spectrum}. As we can see from Fig. \ref{fig_spectrum}, some energies decrease as the applied magnetic field increases. Those energy levels correspond to the $K'$ valley for $m<0$ ($k=-1$ in the formulation of ZZBC). This is an essential behavior that must be contemplated for the calculation of the partition function. That is why, in our numerical calculations, we have considered the azimuthal quantum number $m$  ranging from -50 to 50. 
To guarantee a good convergence in the physical quantities that will be calculated in a range of magnetic field between $0<B\leq 6$ T. In addition, the lowest nonzero electron energy level in ZZBC case (also in the case of IMBC) initially decreases linearly with the magnetic field but then decreases with a Gaussian decay at high magnetic fields.  We use this approximation in our calculations and therefore we fit the energy levels with the function

\begin{figure}[!ht]
\centering
\includegraphics[width=0.9\linewidth]{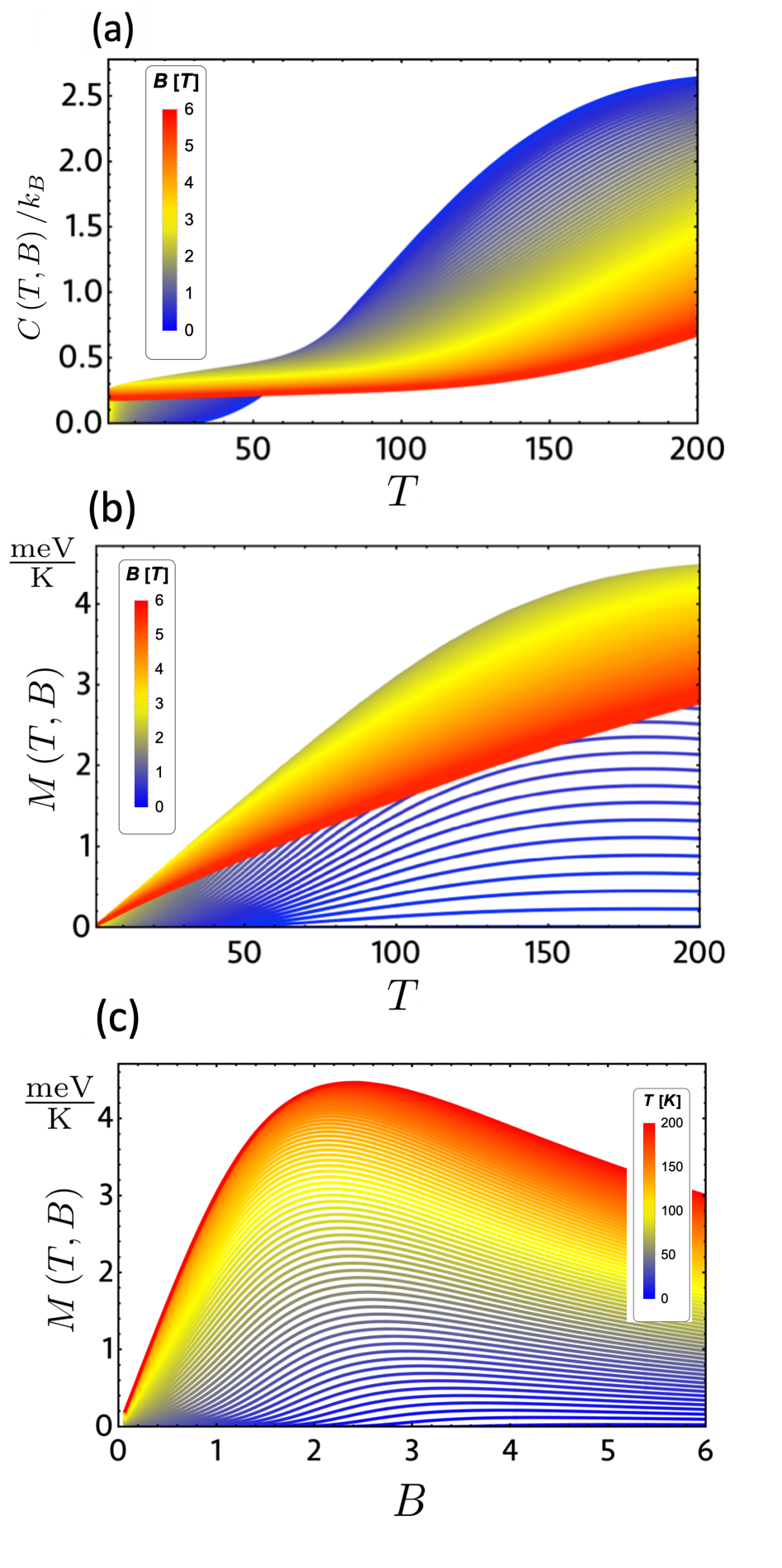}
\caption{(a) Specific heat and (b) magnetization as a function of temperature in the range of 0.1 K to 200 K for different values of the external magnetic field in the range of 0.01 T to 6 T (blue to red color). (c) Magnetization as a function of the external magnetic field for different values of temperatures in the range of 0.1 K to 200 K (blue to red color respectively).}
\label{specific_heat}
\end{figure}

\begin{equation}
    \epsilon (\beta)=a e^{-\left(\frac{\beta-b}{c}\right)^{2}},
\end{equation}
where $a, b$ and $c$ are fitting parameters that depend on the different values of $m<0$ in the $K'$ valley.

The thermodynamic quantities of the system are defined accordingly as
\begin{eqnarray}
\mathcal{F}=-k_{B} T \ln\left[\mathcal{Z}(T,B)\right], \quad S=\left(\frac{\partial \mathcal{F}(T,B)}{\partial T}\right)_{B}
\label{classicalentropy1}
\end{eqnarray}

\begin{eqnarray}
U(T,B)=k_{B}T^{2}\left(\frac{\partial\ln\mathcal{Z}(T,B)}{\partial T}\right)_{B}, 
\label{internalenergy}
\end{eqnarray}

\begin{eqnarray}
C_{B}=\left(\frac{\partial U(T,B)}{\partial T}\right)_{B},
\end{eqnarray}
and

\begin{eqnarray}
M\left(T,B\right)=-\left(\frac{\partial \mathcal{F}}{\partial B}\right)_{T},
\label{magnetiza}
\end{eqnarray}
 
 \begin{figure}[!ht]
\centering
\includegraphics[width=1.0\linewidth]{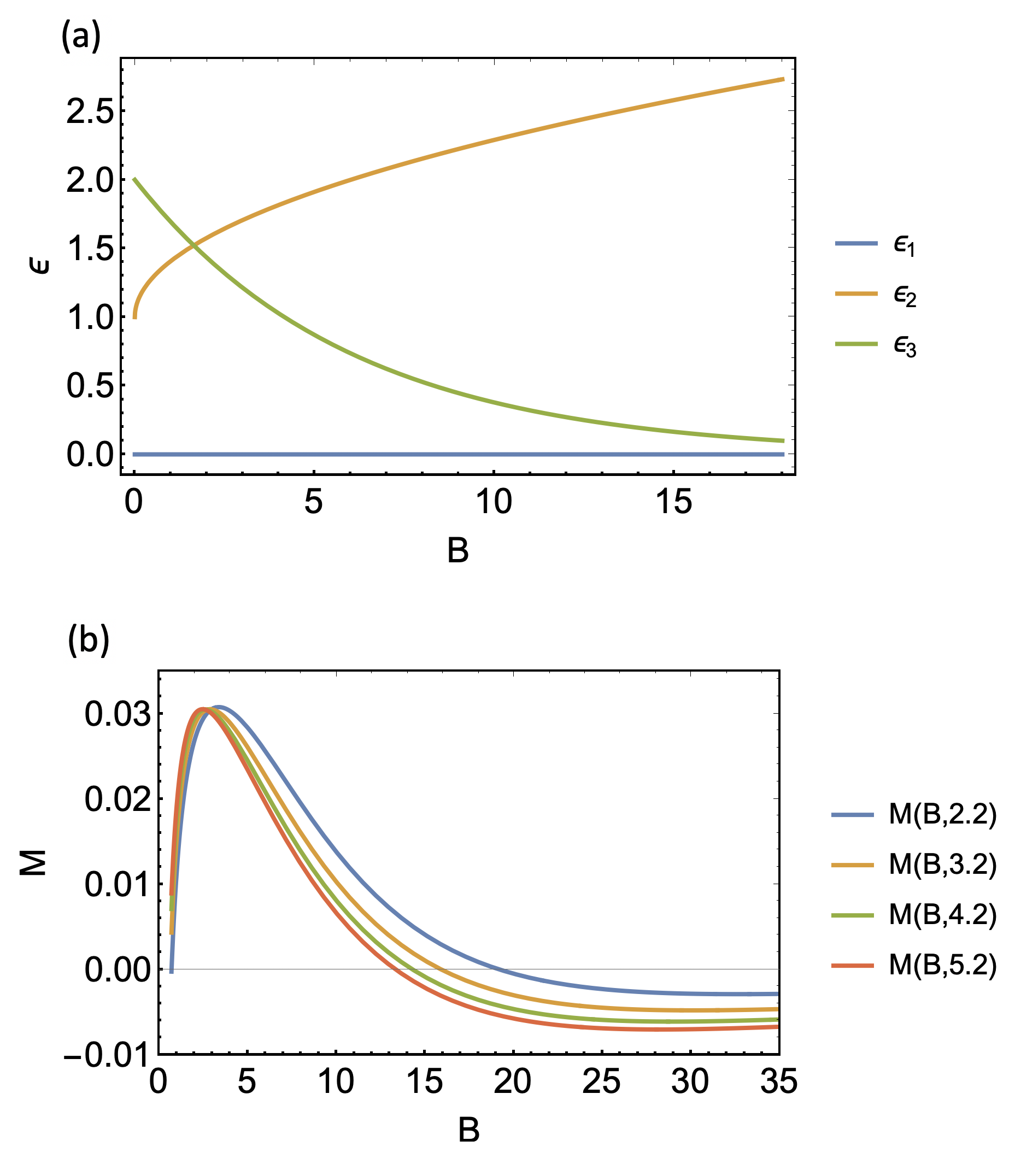}
\caption{(a) Energy levels (in arbitrary units) for the ``toy model" proposed to understand the full numerical results of the graphene quantum dot. The energy states $\epsilon_{1}$, $\epsilon_{2}$ and $\epsilon_{3}$ represent the zero energy state, the Landau level of pristine graphene and the solution for $m$ negatives states of $K'$ point respectively. (b) Magnetization as a function of the magnetic field for different values of the temperature  ranging from $2.2$ up to $5.2$ (in arbitrary units) for the ``toy model" proposed. As we can see for low external magnetic fields, we have positives values of magnetization, while at higher magnetic fields, we have negatives values for $M$.}
\label{modeltoy1}
\end{figure}

where $\mathcal{F}, S, U, C_{B}$ and $M$ are the free energy, entropy, internal energy, specific heat at constant magnetic field and the magnetization of the system, respectively.  In Fig. \ref{specific_heat}(a) we plot the specific heat as a function of temperature and external magnetic field applied. First, we observe that for a temperature range lower than $T \sim 50$ K, magnetic fields under than 2.5 T, the systems has a specific heat smaller than those in the range between 2.6 and 6 T.  Near $T \sim 50$ K there is a change in the behavior of the specific heat for fields lower than 2.5 T, where it is observed that the highest specific heat is obtained at the lowest external magnetic field value applied.  On the other hand, the magnetization as a function of temperature and the external magnetic field is presented in Fig. \ref{specific_heat}(b)  and  in Fig. \ref{specific_heat}(c), respectively. We observe positive values for $M$ in the region between $0<T<200$ K.  As we will see, the magnetization plays a fundamental role in the interpretation of the total work extracted in a cycle whose control parameter is the magnetic field, because the quasi-static work for this cases is given by $W=-\int MdB$. The standard definition of a diamagnetic material is that of a material whose magnetization is negative if the applied magnetic field is positive.  Instead, a material is paramagnetic when the magnetization has the same sign as the applied field. To analyze our magnetization results, we use a simple  three- levels energetic model composed by $\epsilon_{1}=0$, $\epsilon_{2}=1+\sqrt{B/6}$ and $\epsilon_{3}=2e^{-B/6}$ in order to mimic the main feature of the spectrum shown in Fig. \ref{fig_spectrum}. The dimensionless energy spectrum proposed is displayed in Fig. \ref{modeltoy1}(a) where the first energy level simulates the zero-energy state obtained employing ZZBC; the second imitates the Landau levels of pristine graphene and the last energy level $\epsilon_{3}$, deals with the energy levels of $K'$ point for negative values of $m$. If we examine the magnetization (calculated in the same way as Eq. (\ref{magnetiza})) as a function of the magnetic field displayed in Fig. \ref{modeltoy1}(b)  of this ``toy model", we observe that the system have para-and diamagnetic behavior.  The diamagnetic behavior arises because there are branches of the energy spectra that increase with the magnetic field (being Landau levels of graphene which are proportional to $\sqrt{B}$) therefore, they will have a negative magnetization.

\begin{figure}[!ht]
\centering
\includegraphics[width=0.9\linewidth]{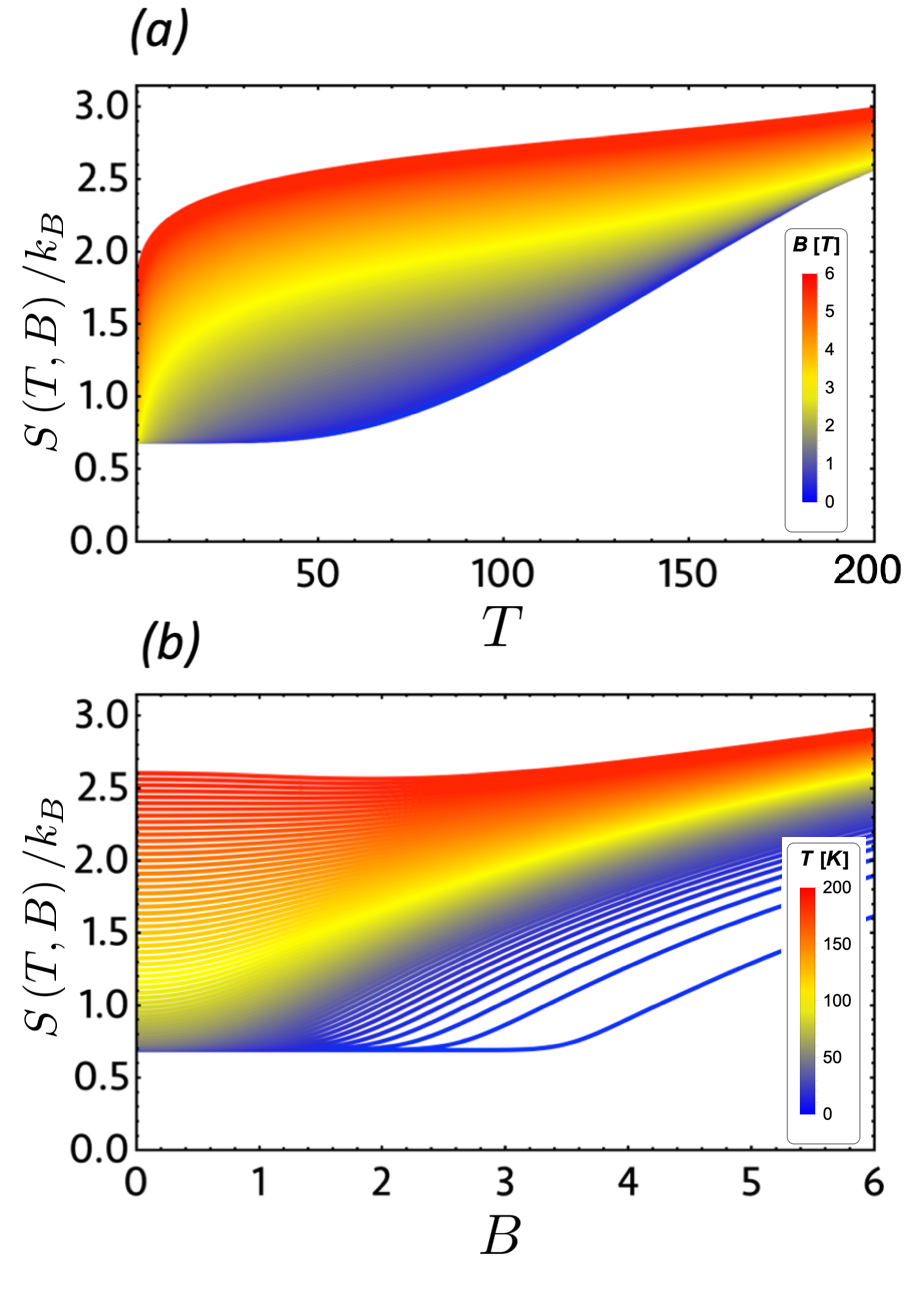}
\caption{(a) Entropy as a function of temperature for different values of the external magnetic field  from 0.1 T to 6 T (blue to red color) and (b) entropy as function of external magnetic field for different values of temperature from $T$=0.1 K to 200 K (blue to red color).}
\label{entropy1}
\end{figure}

On the other hand, a branch of the spectrum whose energy decreases with the magnetic field (the energy levels with negative $m $ for $K'$ point) will have a positive magnetization; consequently, that branch contributes to paramagnetism. Therefore, when both branches are present, both components (para-and diamagnetic) compete, and the one with the more significant probability will prevail. This, of course, will depend on the temperatures and the applied magnetic field. In our real model, we work in a range of temperatures up to 200 K. If we further increase the temperature the populations of higher energy levels start to become relevant in the thermodynamic calculations, and we would thus need to include higher energy levels ($>200$ meV ) breaking the low energy approximation where the continuum approach is valid. Consequently, we only observe a part of the magnetization where the negative $m$ states for the $K'$ point strongly influence.

In Fig. \ref{entropy1}(a), we plot the entropy as a function of temperature, where we see that the entropy is higher as the external field grows. In Fig. \ref{entropy1}(b) we can see the effects of the degeneration of energy levels over $S(T, B)$ for high-temperature behavior and low magnetic field. From Fig. \ref{fig_spectrum}, the energy states present many crossings along the range of $0 < E < 200$ (in meV) for low magnetic field behavior. These crossings are the reason why the entropy for higher temperatures and lower fields tends to collapse to a constant value. On the one hand, as we discussed before, the case of ZZBC exhibits a zero energy state. Therefore, the entropy for $B\rightarrow 0$ and $T\rightarrow 0$ tends also to a constant value as we observe in Fig. \ref{entropy1}(a) and (b) and is proportional to $\ln(2)$ because of the double degeneracy of the zero energy state due to  $K$ and $K'$ valleys. Also, in Fig. \ref{entropy1}(b) we can appreciate a change in the behavior for the entropy in the range of $0<B<1$ (in units of Tesla). This is due to the additional crosses that incorporate the states of $K'$ valley for $m<0$ in that region of the external field. This effect is amplified with temperature because more states are populated that exhibit the above-mentioned crossings.

\section{quantum-adiabatic  and Quasi-static Otto Cycle}

\begin{figure}[!t]
	\centering
	\includegraphics[width=1.0\linewidth]{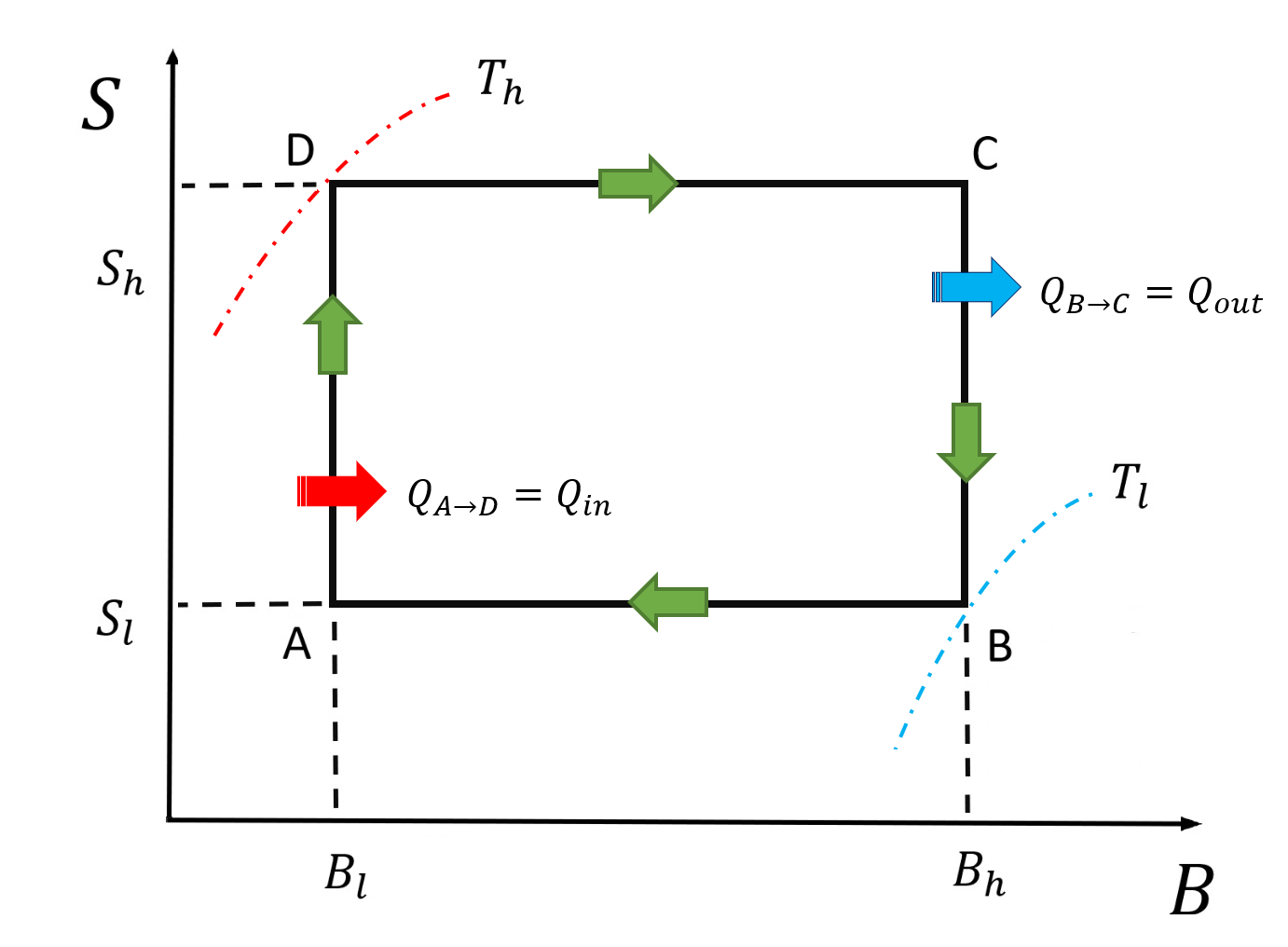}
	\caption{Pictorial description of the proposed Otto cycle.}
	\label{pictorialcycle1}
\end{figure}  

To treat the Otto cycle in the quantum-adiabatic  and quasi-static formulation, we follow the treatment given in Refs. \cite{Quan2005,Pena2015,Pena2016(2)}, which identifies the heat transferred and work performed during a thermodynamic process employing the variation of the internal energy of the system. The quasi-static version of these cycle is composed of four strokes: two isochoric processes and two adiabatic processes. In the quantum version of this cycle, the processes involved are replaced by the respective quantum versions of them. The cycle presented in Fig. \ref{pictorialcycle1} proceeds in the form of $\mathrm{B}\rightarrow \mathrm{A} \rightarrow \mathrm{D}\rightarrow \mathrm{C} \rightarrow \mathrm{B}$, where the processes $\mathrm{A} \rightarrow \mathrm{D}$ and $\mathrm{C} \rightarrow \mathrm{B}$ are isochoric processes while the processes $\mathrm{B}\rightarrow \mathrm{A}$ and $\mathrm{D}\rightarrow \mathrm{C}$ are the adiabatic strokes, respectively. It is important to point out that during the isochoric transformations the system is put in contact with the thermal reservoirs while during the adiabats, the magnetic field is varied. The heat absorbed ($Q_{in}^{q}$) and released ($Q_{out}^{q}$) along the quantum-adiabatic  cycle is given by \cite{Quan2005}

\begin{eqnarray}
\label{qin}
Q_{in}^{q}=\sum_{m}\sum_{\tau}E_{m,\tau}^{l}\left[P_{m,\tau}(T_{h}, B_{l})-P_{m,\tau}^{\mathrm{A}}\right],
\end{eqnarray}

\begin{eqnarray}
\label{qout}
Q_{out}^{q}=\sum_{m}\sum_{\tau}E_{m,\tau}^{h}\left[P_{m,\tau}(T_{l}, B_{h})-P_{m,\tau}^{\mathrm{C}}\right].
\end{eqnarray}
where $T_{h(l)}$ corresponds to the hot (low) reservoir, $E_{m,\tau}^{h,l}$ are the eigenergies of the systems in the quantum isochoric process to an external magnetic field $B_{h}(l)$, $P_{m,\tau}^{\mathrm{A},\mathrm{B},\mathrm{C},\mathrm{D}}$ are the corresponding occupation probabilities along the cycle and the superscript $q$ denotes that is associated to quantum version of the Otto cycle. The net work done in a single cycle can be obtained from $\mathcal{W}^{q}=Q_{in}^{q}+Q_{out}^{q}$, 
\begin{eqnarray}
\label{wtotal}
\mathcal{W}^{q}&=&\sum_{m}\sum_{\tau}\left(E_{m,\tau}^{l}-E_{m,\tau}^{h}\right)\times \\ 
\nonumber
&&\left[P_{m,\tau}(T_{h}, B_{l})-P_{m,\tau}(T_{l}, B_{h})\right],
\end{eqnarray}

The main difference between the quasi-static and quantum-adiabatic  Otto cycle is related to points $A$ and $C$ in the cycle. In the quasi-static case, the working substance can be at thermal equilibrium with a well-defined temperature at each point. On the other hand, for the quantum adibatic case, the working substance only reaches thermal equilibrium in the isochoric stages at points $\mathrm{B}$ and $\mathrm{D}$. After the adiabatic stages, the quantum system is in a diagonal state which is not a thermal state. For the quasi-static engine, the heat absorbed can be calculated by replacing $P^{\mathrm{A}}_{m,\tau}$ with $P(T_{\mathrm{A}}, B_{l})$ in Eq. (\ref{qin}) and the heat released replacing $P^{\mathrm{C}}_{m,\tau}$ with $P(T_{\mathrm{C}}, B_{h})$ in Eq. (\ref{qout}). Therefore, the quasi-static definition of heats involved in the cycle are given by

\begin{equation}
\label{classicalheat1}
Q_{in}^{qs}=U_{\mathrm{D}}(T_{h},B_{l})-U_{\mathrm{A}}(T_{A}, B_{l}), 
\end{equation}

\begin{equation}
\label{classicalheat2}
Q_{out}^{qs}= U_{\mathrm{B}}(T_{l},B_{h})-U_{\mathrm{C}}(T_{C}, B_{h}),
\end{equation}
where $T_{\mathrm{A}}$ and $T_{\mathrm{C}}$ are determined by the condition imposed by the quasi-static isentropic strokes and the superscript $c$ denotes that is associated to quasi-static version of Otto cycle. Therefore, the quasi-static work $(\mathcal{W})$ is given by the difference of four internal energy  in the form \cite{Callen}
\begin{eqnarray}
\label{classicalextre}
\mathcal{W}^{qs}&=& U_{\mathrm{D}}\left(T_{h}, B_{l}\right)-U_{\mathrm{A}}\left(T_{\mathrm{A}}, B_{l}\right)+ \\ 
\nonumber
&& U_{\mathrm{B}}\left(T_{l}, B_{h}\right)-U_{\mathrm{C}}\left(T_{\mathrm{C}}, B_{h}\right).
\end{eqnarray}

Furthermore, the efficiencies are given by
\begin{eqnarray}
\label{effclas}
\eta^{qs}=\frac{\mathcal{W}^{qs}}{Q_{in}^{qs}},
\end{eqnarray}

\begin{eqnarray}
\label{effquan}
\eta^{q}=\frac{\mathcal{W}^{q}}{Q_{in}^{q}}.
\end{eqnarray}

In the case of quasi-static isentropic strokes, we can obtain the intermediate temperatures $(T_{\mathrm{A}}, T_{\mathrm{C}})$ using the entropy function obtaining from Eq. (\ref{classicalentropy1}) and requiring that

\begin{equation}
\begin{split}
S(T_{l},B_{h})=S(T_{\mathrm{A}},B_{l}),  \\
S(T_{h},B_{l})=S(T_{\mathrm{C}},B_{h}).
\end{split}
\end{equation}

For the presentation of efficiency and work results, we define the parameter $r$ given by

\begin{equation}
\label{eqr}
r=\sqrt{\frac{B_{h}}{B_{l}}},
\end{equation} 

that represent the ``compression ratio"  of the problem (in analogy with the case of Otto cycle operating with an ideal gas). Finally, we define the Carnot efficiency as 

\begin{equation}
\eta_{Carnot}=\frac{\Delta T}{T_{h}}=\frac{T_{h}-T_{l}}{T_{h}},
\end{equation}

which serves as a reference value for the efficiency values obtained for this case study.

\section{Results and discussions}

For a correct interpretation of the results for $W$ and $\eta$ that will be shown below, it should be taken into account that given a fixed parameters configuration (i. e. the values of $T_{l}, T_{h}, B_{l}$ and $B_{h}$), a single value of $W$ and $\eta$ is obtained. A black dot will show this particular value over the graphs, and in the left panels of the figures, the corresponding cycle over the thermodynamics quantities is presented. To obtain $W$ and $\eta$ as a function of the $r$ parameter, we fix the values of the isotherms at points D and B (i. e. the values of $T_{h}$ and $T_{l}$ respectively) and the value of the magnetic field at point B (i. e. the value of $ B_ {h}$). The parameter $B_ {l} $ is varied from $B_{h}$ up to an arbitrary minimum value (different from zero and positive) and therefore the parameter $r$ defined in Eq. (\ref{eqr}) varies from one onwards. For the case of the quantum-adiabatic , we only plot the positive work obtained in our calculation.

\subsection{Quasi-static Results}

\label{classicalresults}

\begin{figure}[!ht]
\centering
\includegraphics[width=0.8\linewidth]{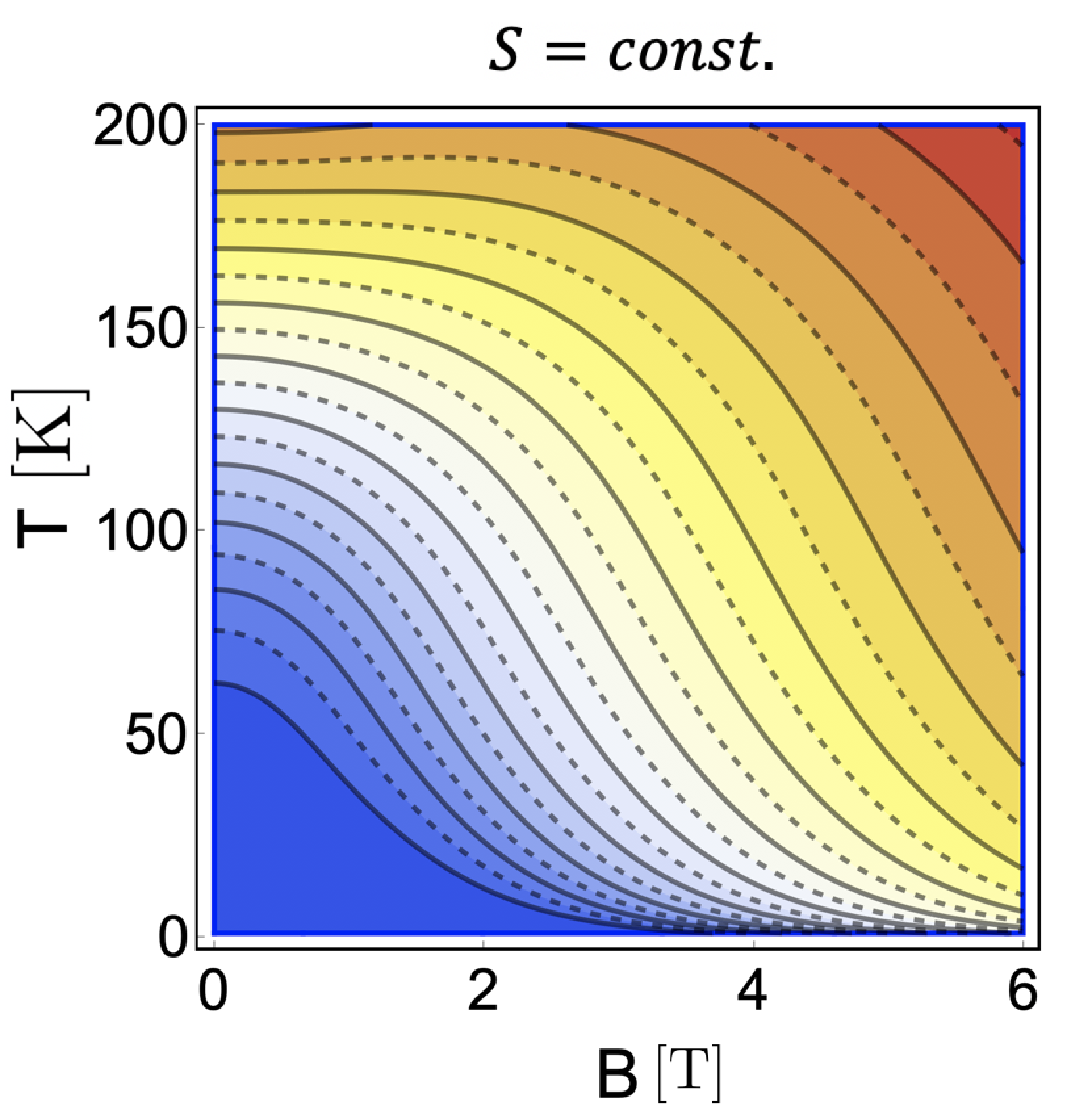}
\caption{The behavior of temperature (vertical axis) versus external magnetic field (horizontal axis) for a quasi-static  isentropic stroke. The contour plot shows the different levels curves (constant entropy values) exhibiting a constant temperature behavior for low magnetic fields. As the field increases, the temperature diminishes to keep the entropy constant.} 
\label{fignueva1}
\end{figure}

If we analyse the condition of constant entropy for the adiabatic stroke, we can obtain the behavior of temperatures and magnetic field along the process as we can appreciate in Fig. \ref{fignueva1} where we observe a decrease in the temperature for an increase in the external magnetic field. This is reflected too, in the way we design cycle proposed in Fig. \ref{pictorialcycle1}, where in our case, lower temperatures are always associated with higher fields and vice versa.

\begin{figure}[!ht]
\centering
\includegraphics[width=1.0\linewidth]{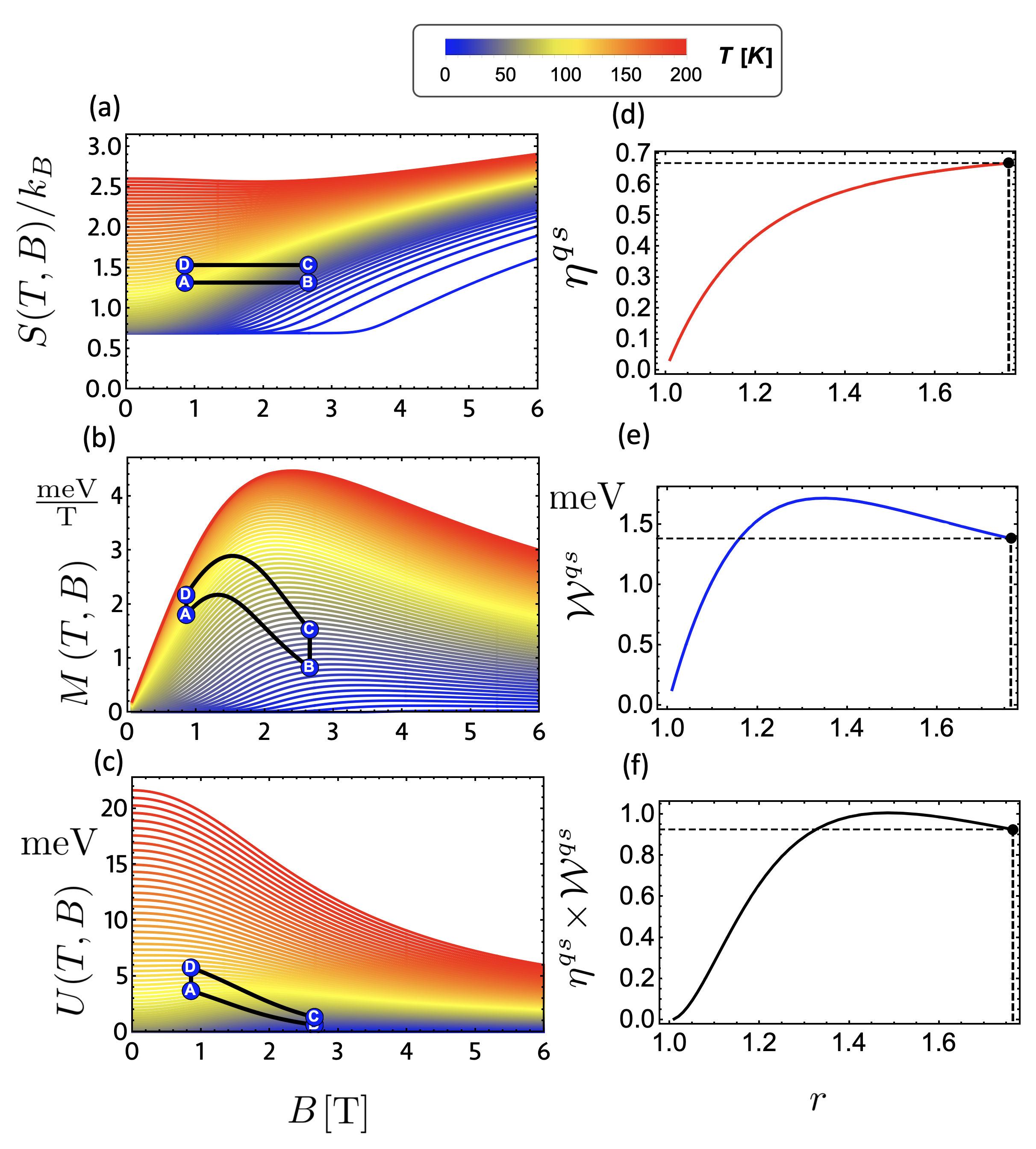}
\caption{Proposed magnetic Otto cycle showing three different thermodynamic  quantities: Entropy~($S$, in units of $k_{B}$), Magnetization ($M$) and Internal Energy ($U$) ((\textbf{a}--\textbf{c}),  respectively) as a function of the external magnetic field and different temperatures from 0.1 K (blue) to 200 K (red).   (\textbf{f}) Efficiency $(\eta^{qs})$ ; (\textbf{e})   the total work extracted $(\mathcal{W}^{qs})$; and (\textbf{f}) the efficiency multiplied by total work extracted $(\eta^{qs} \times \mathcal{W}^{qs})$ for the quasi-static cycle. The black points in  (\textbf{d}--\textbf{f}) represent exactly the cycle B $\rightarrow$ A $\rightarrow$ D $\rightarrow$ C $\rightarrow$ B, presented in  (\textbf{a}--\textbf{c}) panels. The fixed temperatures are $T_{l}=29.9$ K and $T_{h}=119.5$ K and the maximum and minimum values of the external magnetic field are given by $B_{h}=2.65 $ T and $B_{l}=0.85$ T, respectively. Consequently, $r$ moves from 1 to 1.76 approximately.}
\label{fignueva2}
\end{figure}

First, we start with an analysis of a cycle in the central area of the entropy versus external magnetic field diagram. In Fig. \ref{fignueva2} we plot the cycle proposed for the parameters $T_{l}=T_{\mathrm{B}}=29.9$ K, $B_{h}=2.65$ T and $T_{\mathrm{D}}=T_{h}=119.5$ K. However, to maximise the performance keeping $B_h$ constant, we allow  $B_{l}$ to move from 2.65 T to 0.85 T (i. e the compression ratio $r$ moves from 1 to 1.76 approximately). We observe that the maximum value of the total work extracted for this case is given by 1.6 meV  at $r\sim 1.3$ (see Fig. \ref{fignueva2}(e)), which means an optimal value of the maximum external magnetic of $B_h = 1.75$ T,  with and efficiency close to 52$\%$ (see Fig. \ref{fignueva2}(d)). However, this is only the maximum value of total work extracted, and it also matters to see the combination of $\eta$ and $W$ as we can see from Fig. \ref{fignueva2}(f) that indicates that the best configuration is obtained close to $r\sim 1.4$.

\begin{figure}[!ht]
\centering
\includegraphics[width=1.0\linewidth]{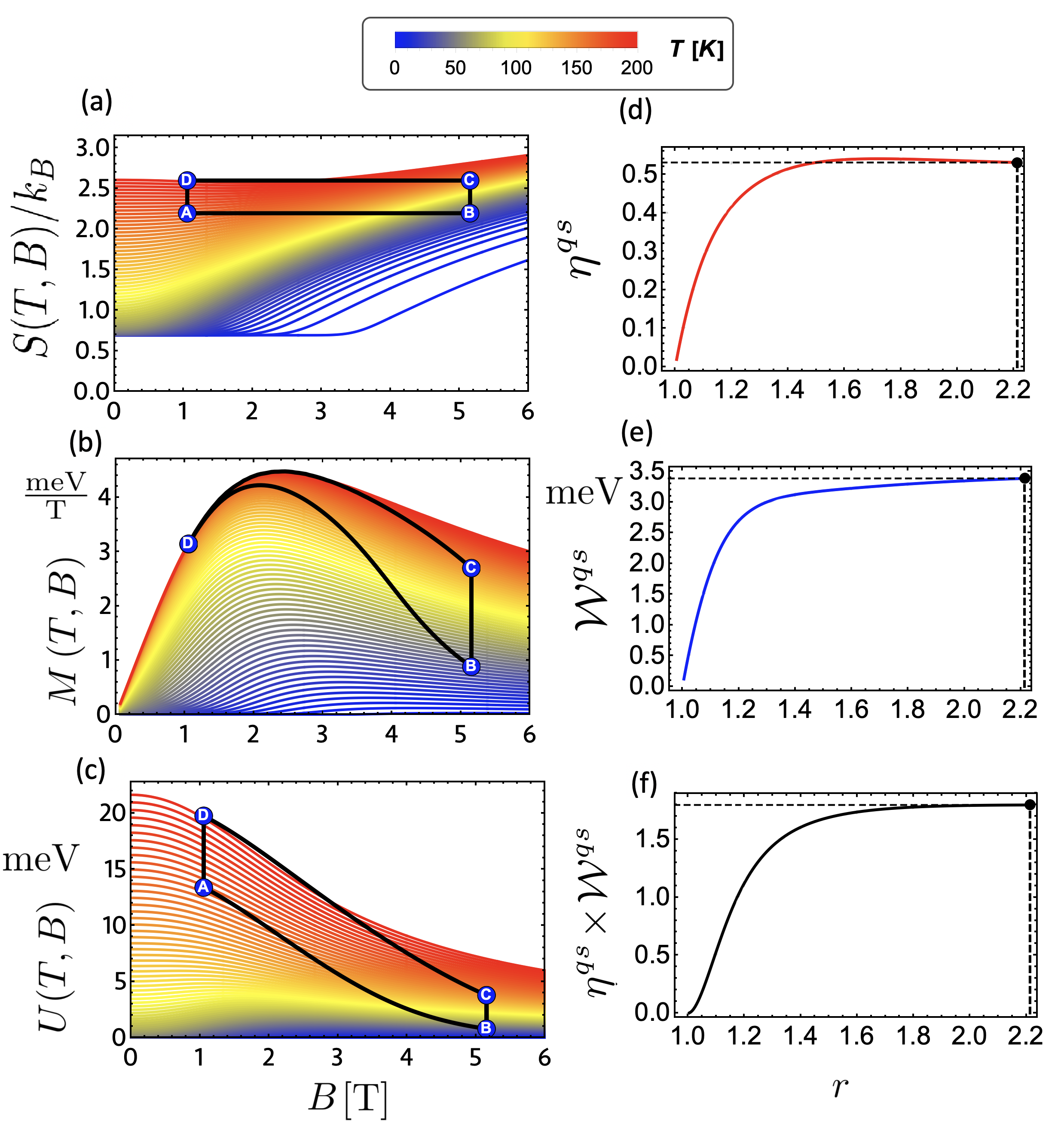}
\caption{Proposed magnetic Otto cycle showing three different thermodynamic  quantities: Entropy~($S$, in units of $k_{B}$), Magnetization ($M$) and Internal Energy ($U$) ((\textbf{a}--\textbf{c}),  respectively) as a function of the external magnetic field and different temperatures from 0.1 K (blue) to 200 K (red), where we observe the effect in the total work extraction due to the collapse of points A and D on the magnetization versus field diagram.(\textbf{d}) Efficiency $(\eta^{qs})$ ; (\textbf{e})   the total work extracted ($\mathcal{W}^{qs}$); and (\textbf{f}) the efficiency multiplied by total work extracted  $(\eta^{qs} \times \mathcal{W}^{qs})$ for the quasi-static cycle. The black points in  (\textbf{d}--\textbf{f}) represent exactly the cycle B $\rightarrow$ A $\rightarrow$ D $\rightarrow$ C $\rightarrow$ B, presented in  (\textbf{a}--\textbf{c}) panels.  The fixed temperatures are $T_{l}=39.15$ K and $T_{h}=200$ K and the maximum and minimum values of the external magnetic field are given by $B_{h}=5.15 $ T and $B_{l}=1.05$ T, respectively. Consequently, $r$ moves from 1 to 2.21 approximately.}
\label{fignueva3}
\end{figure}

On the other hand, a very interesting result in the analysis of $W$ is obtained due to the form of magnetization discussed in Section \ref{thermoqua}. There is the possibility of bringing the points A and D closer in the cycle (over the magnetization diagram), in such a way that a constant work extraction is obtained independent of the change in the external magnetic field (in the range displayed, that means 0 $ <B <$   6 T). This behavior is observed in Fig. \ref{fignueva3}(e), where the maximum value obtained for $W$ is close to 3.4 meV with an efficiency of 50$\%$ for a set of parameters given by $T_{l}=T_{\mathrm{B}}=30.15$ K, $B_{h}=5.15$ T and $T_{\mathrm{D}}=T_{h}=200$ K, and varying $B_{l}$ from 5.15 T to 1.05 T. Consequently, the compression ratio moves from 1 to 2.21 approximately. The explanation for this particular behavior is simply that, quasi-statically, the total work extracted corresponds to the area under the curve of magnetization versus the external magnetic field. Therefore, in Fig. \ref{fignueva3}(b), when approaching points A and D, the contribution of the left-side area begins to be negligible compared to that of the right-hand side, independent of the final $B_{l}$ value over the sample. Therefore, this will cause the work to tend to a constant value, as can be seen in Fig. \ref{fignueva3}(e). It is important to note that this behavior is generated if we made a combination in the parameters in such a way that the temperature of points A and D lies between 150 K and 200 K, where the magnetization has a behavior increasingly close to each other. Also, the maximum value for the efficiency is 51$\%$, and is obtained close to $r\sim 1.7$  and tends to saturate to a value of 50$\%$, 
whose value is below the limit of Carnot efficiency, whose value for this case is $\eta_{\mathrm{Carnot}}\sim 85 \%$.

\begin{figure}[!ht]
\centering
\includegraphics[width=1.0\linewidth]{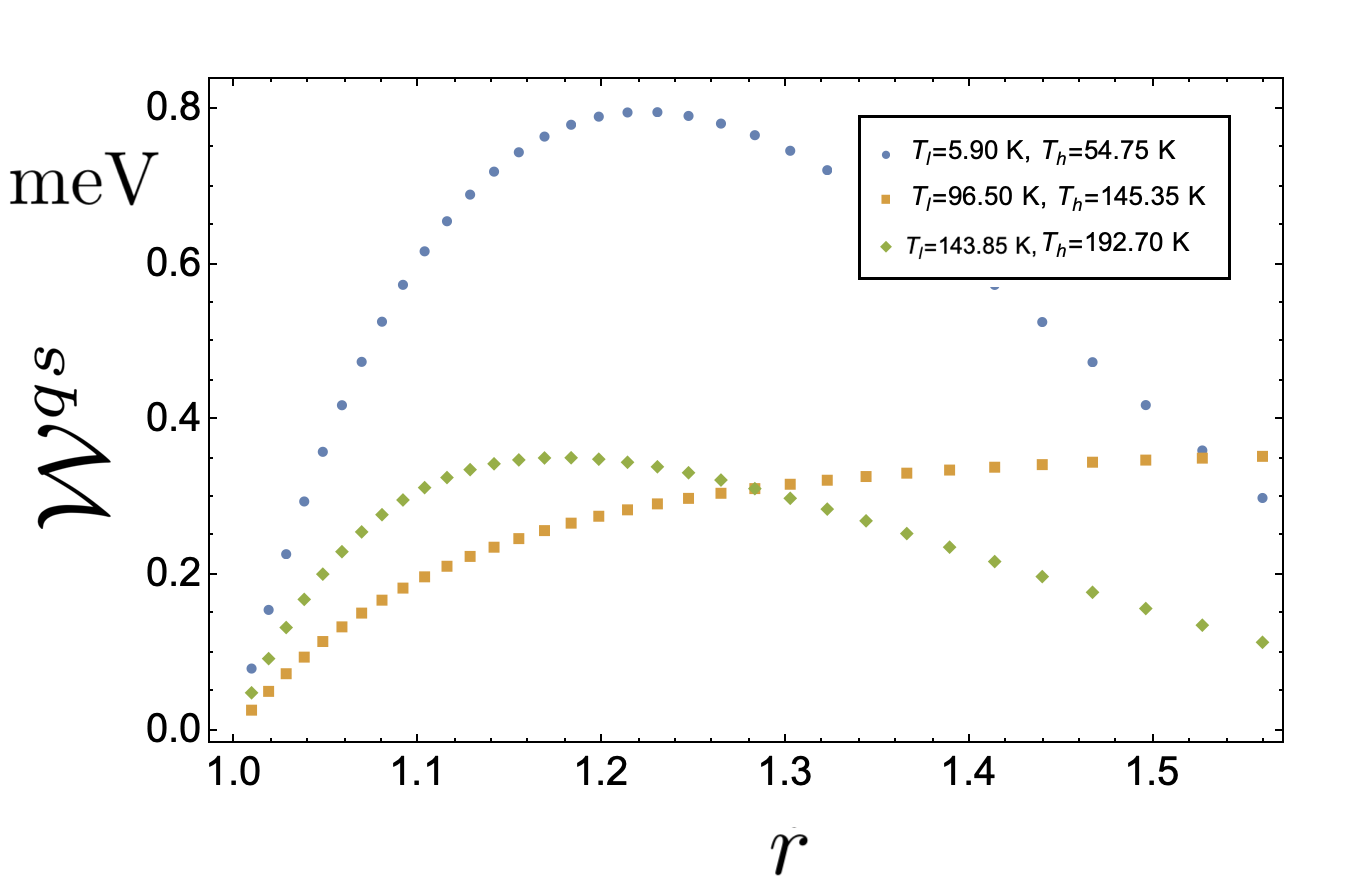}
\caption{Quasi-statically total work extracted $(\mathcal{W}^{c})$ in units of meV as a function of compression ratio for the same values of $\Delta T$ for different regions of temperature. The value of magnetic field are fixed in $B_{h}=2.80$ T and $B_{l}$ moves from 2.80 T to 1.15 T. Therefore the compression ratio $r$ moves from 1 to 1.56.}
\label{fignueva4}
\end{figure}

Finally, we consider working for the same range of the $r$ parameter and maintaining the temperature difference $\Delta T$ in different regions of the entropy versus field diagram considering the cases of low, medium and high temperature to find the best configuration that maximizes the total work extracted. To do that, we fix the value of the external field in $B_{h}=2.80$ T and we move $B_{l}$ from 2.80 T to 1.15 T. Therefore the $r$ parameter moves from 1  to  1.56. We explore three different regions of temperatures given by (i) first zone: $T_{l}=5.90$ K, $T_{h} =54.75$ K, (ii) second zone:  $T_{l}=96.50$ K, $T_{h} =145.35$ K and (iii) third zone: $T_{l}=143.85$ K, $T_{h} =192.70$ K. Our results indicate that we have a better performance for $W$ for low temperature behavior as we can appreciate from Fig. \ref{fignueva4} (circle-dotted line).  As we know, the efficiency associated to a quasi-static engine is always upper bounded by the Carnot efficiency and therefore for the same variation of temperature (i. e. same $\Delta T$), if the temperature of the hot reservoir it is growing ($T_{h}$), lthe corresponding engine efficiency will be smaller. Consequently, a good strategy will be to get a high value of total work for low temperatures, which is observed for our case.

\subsection{Quantum-Adiabatic  Results}

\begin{figure}[!ht]
\centering
\includegraphics[width=0.8\linewidth]{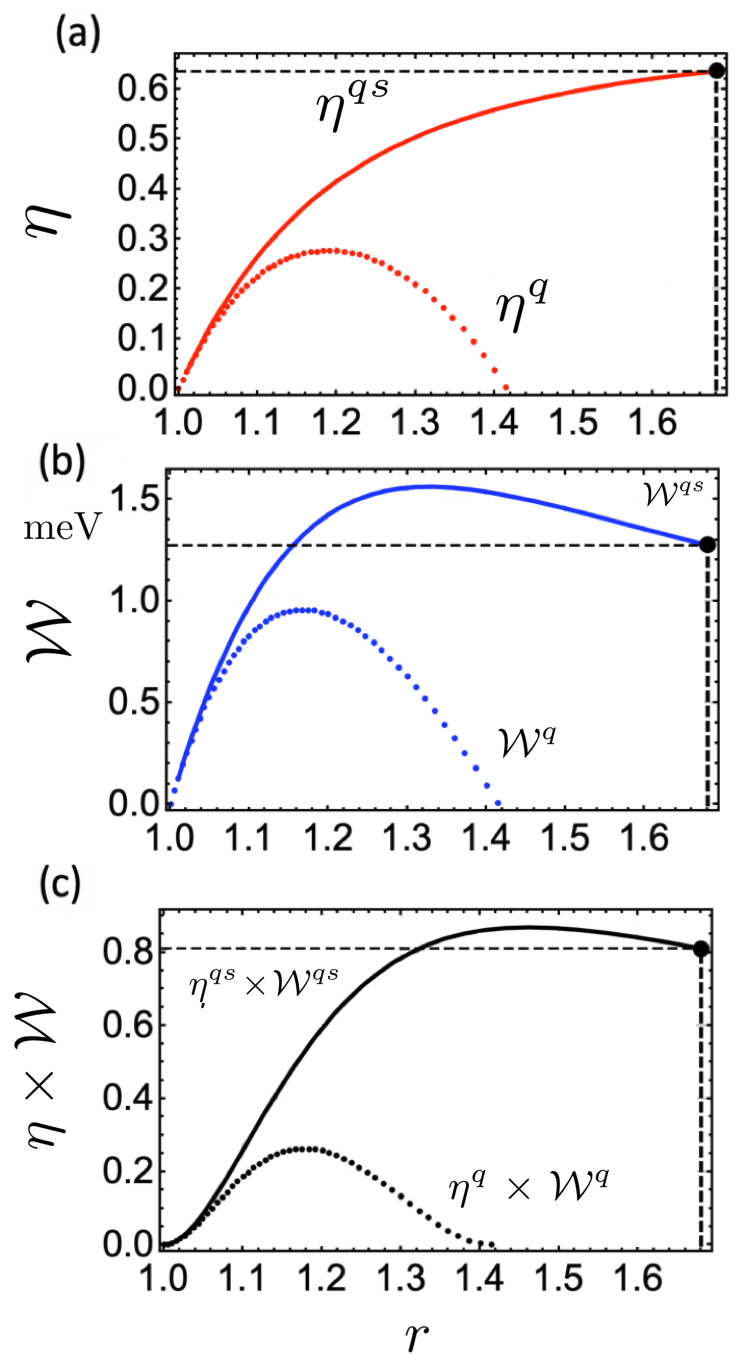}
\caption{(a) Quasi-static ($\eta^{qs}$, solid line) and quantum-adiabatic  ($\eta^{q}$, dotted line) efficiencies , (b) quasi-static ($\mathcal{W}^{qs}$, solid line) and quantum ($\mathcal{W}^{q}$, dotted line) total work extracted and the product of the efficiency by total work extracted for the quasi-static ($\eta^{qs}\times\mathcal{W}^{qs}$, solid line) and quantum ($\eta^{q}\times\mathcal{W}^{q}$, dotted line) cases as a function of the compression ratio $r$ for the same set of parameters of Fig. \ref{fignueva2}. The black point represents exactly the value obtained when we go through the cycle in the form presented in the panels (\textbf{a}--\textbf{c}) of Fig. \ref{fignueva2}.}
\label{figcuantico2}
\end{figure}

Here, we show the results of the evaluation of the quantum-adiabatic  version of the magnetic Otto cycle for the same cases shown in Section \ref{classicalresults}. First, we start from the calculations made with the same parameters of Fig. \ref{fignueva2}. From the panels (a-b) of Fig. \ref{figcuantico2}, we note that the quasi-static and quantum-adiabatic  efficiency and work are equal up to the value of $r \sim 1.07$. This means, for values close to the starting external magnetic field to Point B, we do not notice a difference between the quasi-static and quantum-adiabatic  formulation of the Otto cycle. As shown in Fig. \ref{figcuantico2}(b), we find a transition from positive work to negative work not reflected in the quasi-static scenario close to $ r\sim 1.42$. Additionally, we observe that the maximum value obtained for the quantum-adiabatic  version of $\mathcal{W}$ is noticeably reduced around 0.6 meV compared to its quasi-static counterpart.

\begin{figure}[!ht]
\centering
\includegraphics[width=0.8\linewidth]{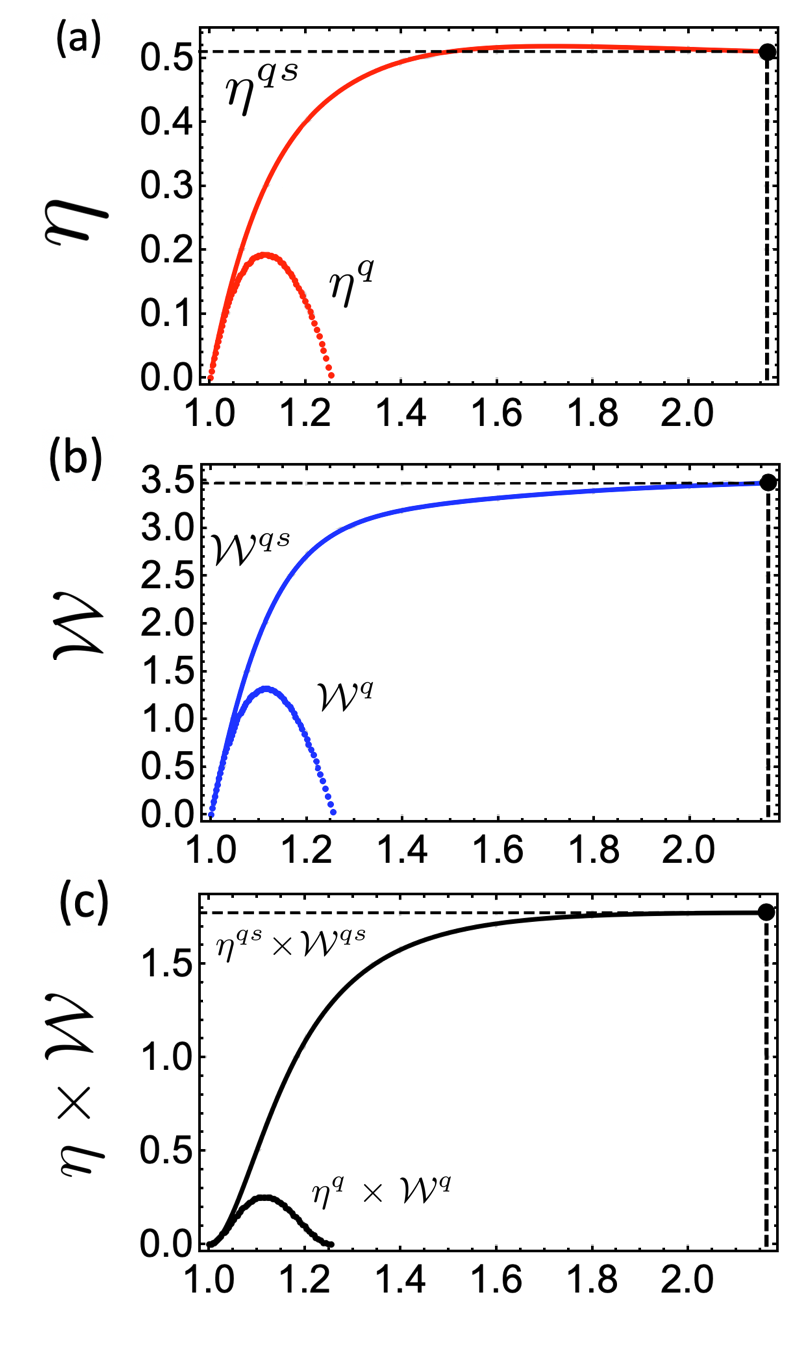}
\caption{(a) Quasi-static ($\eta^{qs}$, solid line) and quantum-adiabatic  ($\eta^{q}$, dotted line) efficiencies , (b) quasi-static ($\mathcal{W}^{qs}$, solid line) and quantum-adiabatic  ($\mathcal{W}^{q}$, dotted line) total work extracted and the product of the efficiency by total work extracted for the quasi-static ($\eta^{qs}\times\mathcal{W}^{qs}$, solid line) and quantum ($\eta^{q}\times\mathcal{W}^{q}$, dotted line) cases as a function of the compression ratio $r$ for the same set of parameters of Fig. \ref{fignueva3}. The black point represents exactly the value obtained when we go through the cycle in the form presented in the panels (\textbf{a}--\textbf{c}) of Fig. \ref{fignueva3}.}
\label{figcuantico3}
\end{figure}

On the other hand, the behavior of constant work extraction and efficiency obtained when we approach the points A and D in the cycle are broken for the quantum formulation of the Otto cycle as we can observe from panels (a-b) of Fig. \ref{figcuantico3}. Moreover, we only observe a faster transition from positive to negative work (close to $r\sim 1.23$), indicating that the machine will operate as a refrigerator rather than as a thermal machine throughout the complete variation of the proposed $r$ parameter. In addition, the maximum value of quantum work for this case dramatically decreases to an amount of the order of 2.5 meV. It is important to remember that we are only plotting positive values of the work and therefore, the efficiency defined by the Eq. (\ref{effquan}) is well defined, and it is bounded to the same range of $r$ where the work obtained is positive.

\begin{figure}[!ht]
\centering
\includegraphics[width=1.0\linewidth]{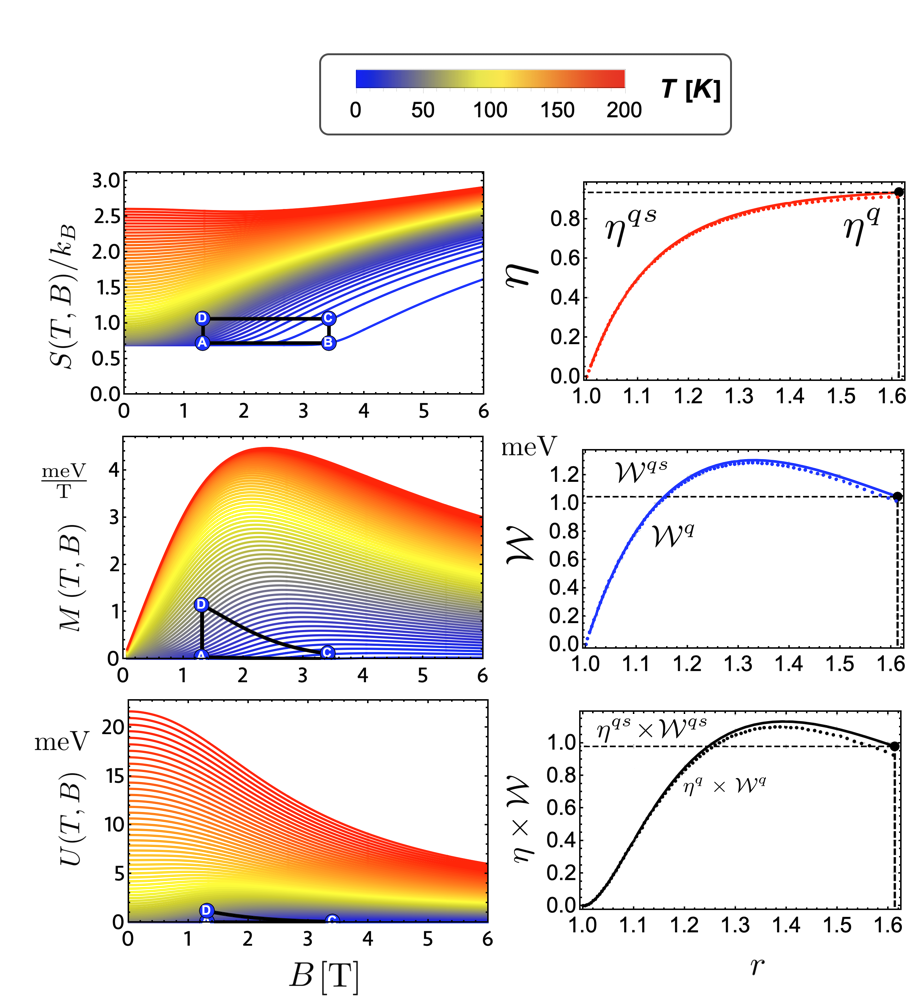}
\caption{Proposed magnetic Otto cycle showing three different thermodynamic  quantities: Entropy~($S$, in units of $k_{B}$), Magnetization ($M$) and Internal Energy ($U$) ((\textbf{a}--\textbf{c}),  respectively) as a function of the external magnetic field and different temperatures from 0.1 K (blue) to 200 K (red). (\textbf{f}) Efficiencies $\eta^{qs}$ and $ \eta^{q}$   (solid and dotted line, respectively); (\textbf{e})   the total works extracted $\mathcal{W}^{qs} $ and $\mathcal{W}^{q}$ (solid and dotted line, respectively); and (\textbf{f}) the efficiency multiplied by their respective total work extracted $ \eta^{qs} \times \mathcal{W}^{qs}$ and $ \eta^{q} \times \mathcal{W}^{q} $  (solid and dotted line, respectively);. The black points in  (\textbf{d}--\textbf{f}) represent exactly the cycle B $\rightarrow$ A $\rightarrow$ D $\rightarrow$ C $\rightarrow$ B, presented in  the panels (\textbf{a}--\textbf{c}).  The fixed temperatures are $T_{l}=1.10$ K and $T_{h}=57.20$ K and the maximum and minimum value of the external magnetic field are given by $B_{h}=3.41 $ T and $B_{l}=1.28$ T, respectively. Consequently, the value of the compression ratio $r$ moves from 1 to  1.63 approximately. }
\label{figcuantico4}
\end{figure}

Finally, our results indicate that in a low temperature regime in the range between 0.1 K to 60 K (i. e. the blue zone over the thermodynamics quantities as a function of the external magnetic field for different temperatures, see Fig. \ref{figcuantico4} ) the quantum-adiabatic  and quasi-static work they are similar to each other in behavior and magnitude as we observe from the left panels of Fig. \ref{figcuantico4} and only note some difference (smaller) between these quantities close to $r\sim 1.3$.

\subsection{Discussions}

Our first result indicates that the quasi-static Otto cycle has a larger total work extracted and efficiency than its quantum-adiabatic  counterpart. To understand this result, we need to remember than in the quasi-static formulation of the Otto cycle the working substance can be in thermal equilibrium at each point in the cycle. Therefore, it is possible to define the temperatures at points A and C in the cycle proposed, and consequently, the internal energy of the systems at these two points can be evaluated. In the quantum-adiabatic case, the working substance is a single system that can only be in a thermal state after thermalizing with the reservoirs, which happens only in the isochoric strokes. Therefore, the points A and C for the quantum case, are diagonal states but not thermal states, thus restricting defining a temperature for said points. If we rewrite the quantum work given by Eq. (\ref{wtotal}) in the form

\begin{eqnarray}
\label{wquant1}
\mathcal{W}^{q}&=&U_{\mathrm{D}}(T_{h},B_{l})+ U_{\mathrm{B}}(T_{l},B_{h}) \\ 
\nonumber
&-&\sum_{m,\tau}\left[E^{l}_{m,\tau}P_{m,\tau}(T_{l},B_{h})+E^{h}_{m,\tau}P_{m,\tau}(T_{h},B_{l})\right],
\end{eqnarray}
where the two first terms appear for the standard definition of the internal energy of the system given for this case by 

\begin{equation}
U=\sum_{m,\tau}E^{l(h)}_{m,\tau}P_{m,\tau}(T_{h(l)},B_{l(h)}).
\end{equation}
The other two terms in the Eq. (\ref{wquant1}) are a type of energy too, but not specifically a thermodynamics definition of $U$, due to the fact that they mix the eigenenergies for low external magnetic field with a probability for the high external magnetic field. If we subtract  Eq. (\ref{wquant1}) , from Eq. (\ref{classicalextre}) we obtain the following equation

\begin{eqnarray}
\label{sustracw}
\mathcal{W}^{qs}-\mathcal{W}^{q}&=&\sum_{m,\tau}E^{l}_{m,\tau}P_{m,\tau}(T_{l},B_{h})- U_{\mathrm{A}}(T_{\mathrm{A}},B_{l}) \\ \nonumber
&+&\sum_{m,\tau}E^{h}_{m,\tau}P_{m,\tau}(T_{h},B_{l})-U_{\mathrm{C}}(T_{\mathrm{C}},B_{h}). 
\end{eqnarray}

The first terms of the last equation correspond to the average of the energy at high magnetic field with thermal probabilities that satisfies the adiabatic condition over the von Neumann entropy in the form

\begin{eqnarray}
\label{entropyq}
S=-k_{B}\sum_{m,\tau}P_{m,\tau}(T_{l},B_{h})\ln\left[P_{m,\tau}(T_{l},B_{h})\right], 
\end{eqnarray}

and correspond to the entropy at point A in the cycle. The internal energy $U_{\mathrm{A}}(T_{\mathrm{A}}, B_{l})$ correspond to the average value of the energy at low magnetic field at temperature $T_{\mathrm{A}}$ with the same value of the entropy present in Eq. (\ref{entropyq}). Therefore, according to principles of thermodynamics \cite{Callen}, $U_{\mathrm{A}}(T_{\mathrm{A}},B_{l})$  is a minimum due to the fact that the entropy given in Eq. (\ref{entropyq}) corresponds to an equilibrium entropy so it must be maximum. Consequently, the quantity $\sum_{m,\tau}E^{l}_{m,\tau}P_{m,\tau}(T_{l},B_{h})$ is always greater or equal to the internal energy $U_{\mathrm{A}}$. The same analysis can be performed for the two final terms of Eq. (\ref{sustracw}), consequently we obtain 

\begin{equation}
\mathcal{W}^{qs}-\mathcal{W}^{q}\geq 0.
\end{equation}
 
 The previous results is general and can be applied to any system where the working substance remains in a diagonal state and does not use quantum resources.
 
At the same time, if we compare the results of the total work extraction in the magnetic Otto cycle for quantum dot modeled by the Fock-Darwin approach and this 2-D system employing the Dirac equation with boundary condition, we note a considerable increase in the total work extraction \cite{Pena2019}. This because the theoretical model of Fock-Darwin consider a parabolic trap that can  be controlled geometrically and is approximately upper-bounded by $\sim 3.0$ meV  for GaAs quantum dots \cite{Kouwenhoven2001}. In particular, the calculation of Ref. \cite{Pena2019} only the value of 1.7 meV is considered due to the fact that the optical transition for cylindrical GaAs quantum dots is approximately around $\sim$ 1 meV for a electrons with effective mass of $0.067 m_{e}$ with $m_{e}$ corresponding to the free electron mass \cite{Jacak,Pacheco2005}. Consequently, the total work extraction for that cases is around $10^{-2}$ meV.  For the case treated in this work, the confinement is imposed by the form of the potential given in Eq. (\ref{potential}) and therefore the energy restriction can only be associated  to the validity of the application of the Dirac equation allowing to work in an energy range of up to 0.2 eV and large dot radii. Accordingly, the total work extraction of this model is greater than the one reported in the Ref. \cite{Pena2019}.
 
 Our second general result indicate that a very low temperature behavior the quantum-adiabatic work and quasi-static work have similar performance. This effect it is observed too for a quantum dot of GaAs in Ref. \cite{Pena2019}, we think that it is a more general concept that can be explained due to the behaviour of thermal populations, and the form of the energy spectrum as a function of magnetic field for these two cases. As we know, at low temperature there are an exponentially decreasing occupation of the higher energy levels. In other words, only the first low lying energy levels define the entropy and energies. On the other hand, by rewriting  Eq. (\ref{sustracw}) in the following form
 
 \begin{eqnarray}
 \label{finaldiscussion}
 \mathcal{W}^{qs}-\mathcal{W}^{q}&=& \sum_{m,\tau}E_{m,\tau}^{l}\left[\frac{e^{-\frac{E_{m,\tau}^{h}}{k_{B}T_{l}}}}{Z(T_{l},B_{h})}-\frac{e^{-\frac{E_{m,\tau}^{l}}{k_{B}T_{\mathrm{A}}}}}{Z(T_{A},B_{l})}\right] \\  \nonumber &+&
 \sum_{m,\tau}E_{m,\tau}^{h}\left[\frac{e^{-\frac{E_{m,\tau}^{l}}{k_{B}T_{h}}}}{Z(T_{h},B_{l})}-\frac{e^{-\frac{E_{m,\tau}^{h}}{k_{B}T_{C}}}}{Z(T_{C},B_{h})}\right], 
 \end{eqnarray}
 
we note that  $\mathcal{W}^{qs}-\mathcal{W}^{q} \rightarrow 0$ when the energies states for the high magnetic field are close in behavior compared to those of low magnetic field and that these states are the predominant ones in the cycle. This is exactly what happens for the structure of the energy spectrum of graphene quantum dots of  Fig. \ref{fig_spectrum}(b) due to the solution obtained for the $K'$ states and the zero-energy state in the ZZBC approximation where these states tend to collapse. Similarly, in the case of quantum dots of GaAs from Ref. \cite{Pena2019}, this behavior in the energy spectrum is obtained due to inclusion of the spin in the model. Additionally, the explanation of why quasi-static work and quantum-adiabatic work obtained for small amounts of the $r$ parameter are equal (as we can see from Fig. \ref{figcuantico4}), is due to the fact that for $r$ close to one, the difference between the temperatures $T_{l}$ with $T_{\mathrm{A}}$ and $T_{h}$ with $T_{\mathrm{C}}$ are tiny and if we additionally add the aforementioned behavior in the energy spectrum, we obtain that the difference between these two values of works is close to zero.
  
  We strongly believe that our approach for this proposal can be further improved. First, concerning of the limitation of the size of the material, the use of the tight-binding approach would allow seeing the effects of size and edge in this system. In addition, the density of states can be calculated, and all the thermodynamics can be recalculated, considering the effects of valence and conduction electrons. Lastly, this work can be extended by employing quantum optimal control or shortcuts to adiabaticity techniques  \cite{Odelin,Werschnik,Glaser} in order to realize experimentally a high-performing Otto cycle.

\section{Conclusions}

In this work, we explored the quasi-static and quantum Otto cycle for the case of a working substance corresponding to a quantum dot of graphene modeled by the Dirac equation with the use of zigzag boundary condition.  We analyzed all the relevant thermodynamics quantities of the system and found that the entropy for low magnetic field tends to a constant value. Also, due to the strong degeneracy of the energy spectrum, the entropy grows along with the external magnetic field for all temperatures considered. In the quasi-static approach, we obtained a region of parameters where the efficiency and total work extracted becomes constant and is not present in the quantum-adiabatic  approach. Moreover, in the quantum case, for that cases, we observe a sudden transition from positive to negative work extraction indicating that the cycle proposed corresponds to a refrigeration cycle rather than a heat engine. Also, we reported a smaller work extraction for the quantum-adiabatic case compared to the quasi-static approach because in the former case the system only thermalizes in the isochoric stages while for the latter case the system goes through four equilibrium states. Hence, because of the principle of minimum energy, the system is allowed to extract more energy when the adiabatic strokes can lead to states that are in thermal equilibrium, which is only possible in the quasi-static case. We recall that in our formulation the working substance remains in a diagonal state and we do not use quantum resources (for example quantum coherence), which in some cases could lead to and enhanced performance.


\section*{Acknowledgements}

F. J. P acknowledges the financial support of CONICYT + PAI CONVOCATORIA NACIONAL SUBVENCI\'ON A INSTALACI\'ON EN LA ACADEMIA CONVOCATORIA 2018 PAI77180015. P. Vargas and O. N. acknowledge support from Financiamiento Basal para Centros Cient\'ificos y Tecnol\'ogicos de Excelencia, under Project No. FB 0807 (Chile). O. N. acknowledge support from USM-DGIIP for Ph.D. fellowship. GDC acknowledges financial support from the UK EPSRC (EP/S02994X/1) .


\bibliographystyle{apsrev}

\end{document}